\documentclass[fleqn,twoside]{article}
\usepackage{espcrc2} 
\usepackage{psfig}

\title{Baryon magnetic moments in  the external field method
\thanks{presented by F.X. Lee at Lattice 2004.}}

\author{ F.X. Lee\address[GW]{Center for Nuclear Studies, 
Department of Physics,
The George Washington University, Washington, DC 20052, USA}, 
R. Kelly\addressmark[GW],
L. Zhou\addressmark[GW],
W. Wilcox\address{Department of Physics, 
Baylor University, Waco, TX 76798, USA}
}

\begin{document}

\begin{abstract}
We present a calculation of the magnetic moments of the baryon octet and decuplet
using the external field method and standard Wilson gauge and fermion actions in the quenched approximation.
Progressively smaller static magnetic fields are introduced on a $24^4$ lattice
at beta=6.0 and the pion mass is probed down to about 500 MeV.
Magnetic moments are extracted from the linear response of the masses to the external field.
\end{abstract}

\maketitle

 \section{Method}
Magnetic moments are an important fundamental property of particles. They
determine dynamical response of a bound system to external perturbations, 
and provide valuable insight into internal strong interaction structure. 
There are at least two ways to extract the magnetic moment. 
One is from form factors (three-point functions)
where an extrapolation to $G_M(Q^2=0)$ is required. 
The other is the external field method using only two-point 
functions~\cite{mart82,bernard82,smit87,rubin95}.
Here we report a calculation in the latter. It is an extension of our work 
on electric and magnetic polarizabilities~\cite{joe04,zhou02,zhou04}.

In order to place a magnetic field on the lattice, we construct an 
analogy to the continuum case. The fermion action is modified 
by the minimal coupling prescription 
$D_\mu = \partial_\mu+gG_\mu + q A_\mu$
where $q$ is the charge of the fermion field and $A_\mu$ is the vector 
potential describing the external field. On the lattice, the prescription
amounts to a modified link variable $U_\mu^\prime=U_\mu U_\mu^{(B)}$.
Choosing $A_y = B x $, a constant magnetic field B can be introduced 
in the $z$-direction.  Then the phase factor is 
$
U_\mu^{(B)}=\exp{(iqa^2Bx)}.
$
On a finite lattice with periodic boundary conditions, to get a constant magnetic field, 
B has to be quantized by the condition
$
qBa^2={2\pi n \over N_xN_y}, \hspace{1mm} n=1,2,3,\cdots
$
to ensure that the magnetic flux through plaquettes in the xy-plane is contant.
However, for $N_x=N_y=16$ and $1/a=2 GeV$, the lowest field would give the proton 
a mass shift of about 390 MeV, which is too large (the proton is severely distorted). 
We abandon the quantization condition and choose to work with smaller fields.
To minimize the boundary effects, we work with fixed (or Dirichlet) b.c. in the x-direction and 
large $N_x$, so that quarks originating in the middle of the lattice have little chance of propagating to the edge.

We use $24^4$ lattice at $\beta=6.0$, and six kappa values $\kappa$=0.1515, 0.1535, 0.1545, 
0.1555, corresponding to pion mass of 1015, 908, 794, 732, 667, 522 MeV.
The strange quark mass is set at $\kappa$=0.1535. The souce location is (x,y,z,t)=(12,1,1,2).
We analyzed 87 configurations.
The following five dimensionless numbers 
$\eta=qBa^2$=+0.00036, -0.00072, +0.00144, -0.00288, +0.00576 give four small B fields at 
$eBa^2$=-0.00108, +0.00216,  -0.00432, +0.00864 for both u and d (or s) quarks. 
Note that "small" here is in the sense that the mass shift is only a fraction of the proton mass:
$\mu B/m \sim$ 0.6 to 4.8\% at the smallest pion mass. 
In absolute terms, $B\sim 10^{13}$ tesla.

The mass shift in the presence of small fields can be expanded as
a polynomial in B,
\begin{equation}  \label{mshift}
\delta m (B) = m(B)-m(0)= c_1 B + c_2 B^2+ c_3 B^3+\cdots
\end{equation}
To eliminate the contamination from the even-power terms, we calculate mass shifts 
both in the field $B$ and its reverse $-B$, then take the difference and divide by 2.
So our calcaulation is equivalent to 11 separate spectrum calculations: 5 orignal $\eta$ values, 
5 reversed, plus the one at zero field to set the baseline.
For a Dirac particle of spin $s$ in uniform fields, $E_\pm=m_\pm\pm\mu B$ 
where the upper sign means spin up and the lower sign means spin-down, and 
$\mu=g {e\over 2m}s$. We tried the following three methods to 
extract the g factors, and found they are equivalent within statistical errors.
\begin{equation}
g=\left( {2m_+ m_- \over m_+ + m_-} \right){(E_+ - m_+)-(E_- - m_-)\over eBs}
\label{g1}
\end{equation}
\begin{equation}
g={m_+ (E_+ - m_+) - m_- (E_- - m_-)\over eBs}
\end{equation}
\begin{equation}
g={(E_+^2 - m_+^2) - (E_-^2 - m_-^2)\over 2eBs}
\end{equation}
The results quoted are from Eq.~\ref{g1}.

 \section{Results and discussion}
\begin{figure}
\centerline{\psfig{file=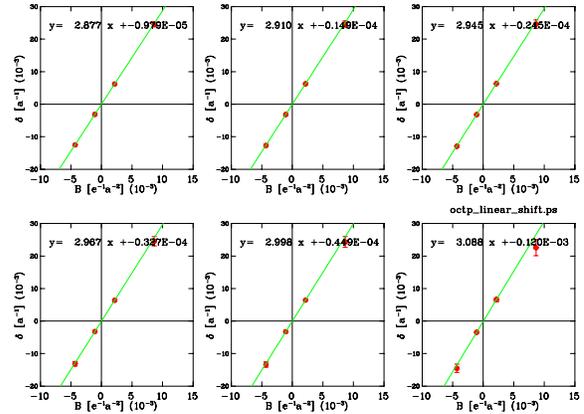,width=7.5cm,angle=90}}
\vspace*{-0.9cm}
\caption{Proton mass shifts as a function of magnetic field 
$B$ at the six quark masses. The line is a fit using only the two smallest B values.}
\label{octp_linear_shift}
\vspace*{-0.5cm}
\end{figure}

\begin{figure}
\centerline{\psfig{file=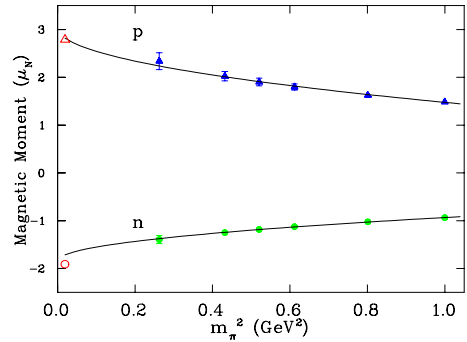,width=6.0cm,angle=90}}
\vspace*{-0.9cm}
\caption{Magnetic moments (in nuclear magnetons) for the proton and neutron 
as a function of $m_\pi^2$. The line is a simple chiral fit.
The experimental values, taken from the PDG~\protect\cite{pdg04}, 
are indicated by the empty symbols.}
\label{mag_pn}
\vspace*{-0.5cm}
\end{figure}

\begin{figure}
\centerline{\psfig{file=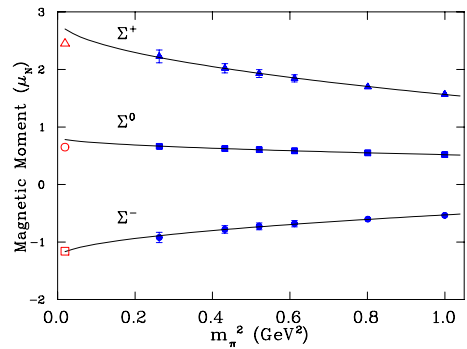,width=6.0cm,angle=90}}
\vspace*{-0.9cm}
\caption{Magnetic moments for the octet $\Sigma$.}
\label{mag_osig}
\vspace*{-0.5cm}
\end{figure}

\begin{figure}
\centerline{\psfig{file=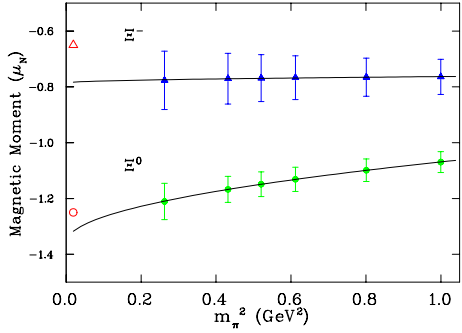,width=6.0cm,angle=90}}
\vspace*{-0.9cm}
\caption{Magnetic moments for the octet $\Xi$.}
\label{mag_oxi}
\vspace*{-0.4cm}
\end{figure}

\begin{figure}
\centerline{\psfig{file=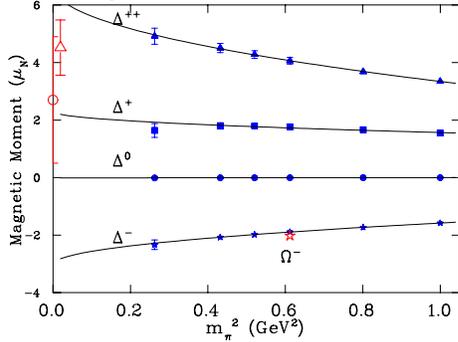,width=6.0cm,angle=90}}
\vspace*{-0.9cm}
\caption{Magnetic moments for the $\Delta$.}
\label{mag_decd}
\vspace*{-0.3cm}
\end{figure}

\begin{figure}
 \centerline{\psfig{file=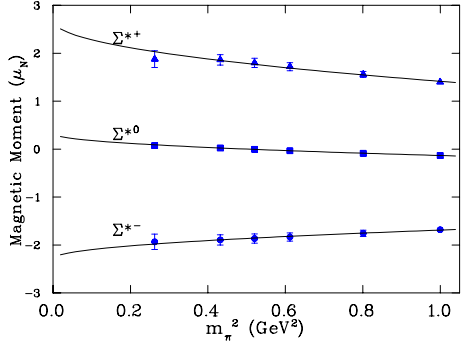,width=6.0cm,angle=90}}
 \vspace*{-0.9cm}
 \caption{Magnetic moments for the decuplet $\Sigma^*$.}
 \label{mag_dsig}
 \vspace*{-0.3cm}
 \end{figure}

\begin{figure}
 \centerline{\psfig{file=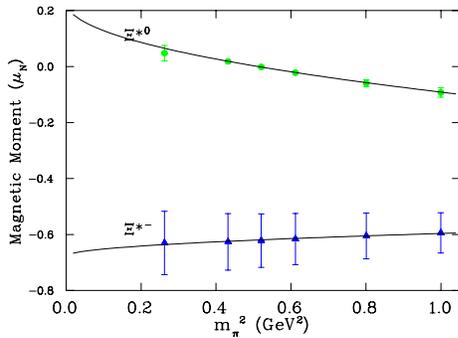,width=6.0cm,angle=90}}
 \vspace*{-0.9cm}
 \caption{Magnetic moments for the decuplet $\Xi^*$.}
 \label{mag_dxi}
 \vspace*{-0.3cm}
 \end{figure}

\begin{figure}
\centerline{\psfig{file=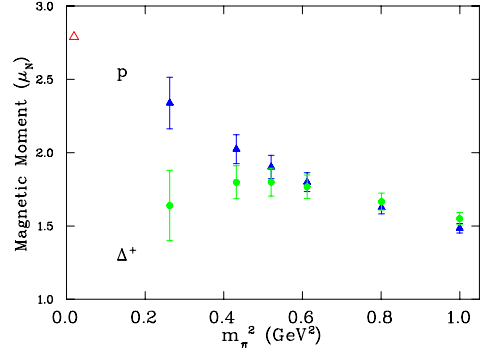,width=6.2cm,angle=90}}
\vspace*{-0.9cm}
\caption{Comparison of magnetic moments of the proton and $\Delta^+$.}
\label{mag_pdecdp}
\vspace*{-0.3cm}
\end{figure}

Fig.~\ref{octp_linear_shift} displays a typical mass shift as a function of the field 
for the proton.
There is good linear behavior going through the origin, 
an indication that contamination from the higher terms has been 
effectively eliminated. We only use the two smallest field values 
to do the linear fit.
Fig.~\ref{mag_pn} shows the results for the proton and the neutron.
The line is a simple chiral fit using the ansatz
$
\mu=a_0 + a_1 m_\pi.
$
A more advanced chiral extrapolation could be carried out,
using, for example,  the prescription in Ref.~\cite{young04,leinweber04}.
Fig.~\ref{mag_osig} to Fig.~\ref{mag_dxi} show the results for other particles.
The agreement with experiment for the $\Omega^-$ is encouraging 
since no extrapolation is needed for this point.
The experimental value for $\Delta^+$ (slightly shifted for better view) is 
taken from Ref.~\cite{kotulla02}. 
Finally, Fig.~\ref{mag_pdecdp} shows the proton and $\Delta^+$ together.
The opposite curvatures are a signature of quenched chiral physics.
A similar behavior has been observed by ~\cite{zanotti03,young03} using 
the form factor method.
%
%
%
%
%
%
%
%
%
%
%
%
%
%
%
%
%
%


This work is supported in part by U.S. Department of Energy
under grant DE-FG02-95ER40907, and by NSF grant 0070836.
The computing resources at NERSC and JLab have been used.

\end{document}